\documentstyle[aaspp4,11pt,psfig]{article}
\tightenlines

\def\dw#1{}			
\def\jm#1{}

\newcommand{\lya}{Ly$\alpha$\ }
\newcommand{\no}{log$N({\rm HI})$\ }
\newcommand{\nhi}{N_{\rm HI} }

\newcommand{\kms}{\, {\rm km \, s}^{-1} }
\newcommand{\cm}{{\, \rm cm} }
\newcommand{\mpc}{\, {\rm Mpc} }
\newcommand{\etal}{et al.\ }

\newcommand{\ergs}{erg cm$^{-2}$ s$^{-1}$ sr$^{-1}$ Hz$^{-1}$\ }

\begin{document}


\title{The Opacity of the Lyman
Alpha Forest and Implications for $\Omega_{baryon}$ and the Ionizing Background\altaffilmark{1}}

\vskip 1.5cm

\author{Michael Rauch\altaffilmark{2,10}, 
Jordi Miralda-Escud\'e\altaffilmark{3,4}, 
Wallace L.W. Sargent\altaffilmark{2}, Tom A. Barlow\altaffilmark{2},
David H. Weinberg\altaffilmark{5}, Lars Hernquist\altaffilmark{6,11},
Neal Katz\altaffilmark{7,8}, Renyue Cen\altaffilmark{9},  
Jeremiah P. Ostriker\altaffilmark{9}}

\altaffiltext{1}{The observations were made at the W.M. Keck Observatory
which is operated as a scientific partnership between the California
Institute of Technology and the University of California; it was made
possible by the generous support of the W.M. Keck Foundation.}

\altaffiltext{2}{Astronomy Department, California Institute of Technology,
Pasadena, CA 91125, USA}
\altaffiltext{3}{Institute for Advanced Study, Princeton, NJ 08540}
\altaffiltext{4}{Department of Physics and Astronomy,
University of Pennsylvania, Philadelphia, PA 19104 (present address)}
\altaffiltext{5}{Department of Astronomy, The Ohio State University, Columbus, OH 43210}
\altaffiltext{6}{Lick Observatory, University of California, Santa Cruz, CA 95064}
\altaffiltext{7}{Department of Astronomy, University of Washington, Seattle, WA 98195}
\altaffiltext{8}{Department of Physics and Astronomy, University of Massachusetts, Amherst,
MA, 98195}
\altaffiltext{9}{Princeton University Observatory, Princeton, NJ 08544}
\altaffiltext{10}{Hubble Fellow}
\altaffiltext{11}{Presidential Faculty Fellow}

\medskip 
\vskip 2.5cm

{Subject Headings:  cosmology: observation ---
intergalactic medium --- quasars: absorption lines} 

\vskip 3.0cm
\centerline{\it submitted to the Astrophysical Journal}
\vskip 2.0cm


\vfill

\pagebreak

\begin{abstract}
We have measured the distribution function of the flux decrement
$D=e^{-\tau}$ caused by \lya forest absorption from intervening gas in
the lines of sight to high redshift QSOs from a sample of seven high
resolution QSO spectra obtained with the Keck telescope. The observed
flux decrement distribution function (FDDF) is compared to the FDDF from
two simulations of the \lya forest: a $\Lambda$CDM model
(with $\Omega$ =0.4, $\Lambda$=0.6) computed with the Eulerian code of
Cen \& Ostriker, and a standard CDM model
(SCDM, with $\Omega$ = 1) computed with the SPH code of
Hernquist, Katz, \& Weinberg.
Good agreement is obtained between the shapes of the simulated and
observed FDDFs for both simulations after fitting only one free
parameter, which controls the mean flux decrement.
The difference between the predicted FDDFs from the two simulations is
small, and we show that it arises mostly from
a different temperature in the low-density gas
(caused by different assumptions that were made about
the reionization history in the two simulations), rather than 
differences between the two cosmological models {\it per se}, or
numerical effects in the two codes which use very different computational
methods. 

  A measurement of the parameter $\mu \propto \Omega^2_{b}h^{3}
 / \Gamma$ (where
$\Gamma$ is the HI ionization rate due to the ionizing background) is
obtained by requiring the mean flux decrement in the simulations to
agree with the observed one.  Estimating the lower limit $\Gamma$ $>
7\times10^{-13}$ s$^{-1}$ from the abundance of known QSOs, we derive a
lower limit on the baryonic matter density, $\Omega_{b}
h^{2}$$>0.021$($0.017$) for the $\Lambda$CDM (SCDM) model. The difference
between the lower limit inferred from the two models is again due to 
different temperatures in the low-density gas. We give general analytical
arguments for why this lower limit is unlikely to be reduced for any
other models of structure formation by gravitational collapse that can
explain the observed \lya forest.
The large $\Omega_{b}$ we infer is inconsistent with some recent D/H
determinations (Rugers \& Hogan 1996a,b), favoring a low deuterium
abundance as reported by Tytler, Fan \& Burles (1996). Adopting a
fixed $\Omega_{b}$, the measurement of $\mu (z)$ allows a determination
of the evolution of the ionizing radiation field with redshift. Our
models predict an intensity that is approximately constant with
redshift, which is in agreement with the assumption that the ionizing
background is produced by known quasars for $z < 3$, but requires
additional sources of ionizing photons at higher redshift given the
observed rapid decline of the quasar abundance.

\end{abstract}

\pagebreak

\section{Introduction}

The mean baryon density of the universe is one of the observationally
most relevant cosmological parameters. It influences the whole range of
observable baryonic structures, from the abundances of primordial
nuclei to the observational appearance of the large-scale distribution
of intergalactic gas and galaxies.  At the same time, the value of 
$\Omega_b$ is a testable prediction of the Standard Big Bang model.
Using the theory of primordial nucleosynthesis in the early universe we
can compute the abundances for the primordial gas as a function of only
one parameter, the cosmic baryon density. Consequently, $\Omega_b$ can be
inferred from the measured abundances of the light elements (e.g.,
Walker \etal 1991).  The agreement of the various abundances with
observations has long been held as one of two most important successes
of the Big Bang theory (the other one being the prediction of the
CMB).  Recently, attempts have been made to measure $\Omega_b$ from the
deuterium abundance in high-redshift QSO absorption systems (e.g.,
Songaila et al. 1994; Carswell et al. 1994; Tytler, Fan, \& Burles
1996; Burles \& Tytler 1996; Rugers \& Hogan 1996a,b).  These
observations exploit the fact that the D/H ratio is highly sensitive to
the baryon density. The absorbing gas clouds have low metallicity, so
the deuterium abundance should reflect the primordial value and not be
excessively affected by stellar processing. The interpretations of
these observations are currently subject to discussion and at the time
of writing there is no agreement among different groups about the value
of the deuterium abundance.

The predictions of primordial nucleosynthesis would be much more
impressive if there were an {\it independent} method to measure
$\Omega_b$ and it was found to agree with the value required by the
abundances of the light elements.   In the long run, the best independent
measurements may come from
observations of the spectrum of fluctuations in the microwave
background, which can yield $\Omega_b$ from the amplitude of the peak
caused by acoustic waves (Holtzman 1989; Jungman \etal 1995), 
as well as other cosmological parameters.

Another possible approach involves counting baryons more ``directly''
by adding up the contribution to the mean cosmic density from all classes
of known astronomical objects. Until now such work has been limited to
low redshift objects, such as galaxies and galaxy clusters (e.g. Persic \&
Salucci 1992; Bristow \& Phillips 1994). An unknown fraction
of the dark matter known to exist in galaxy halos and clusters
of galaxies could be of baryonic origin, and baryons may also hide in
the intergalactic medium or in low surface brightness galaxies,
where even their gravitational influence is hard to detect.
Thus, only a lower limit to the baryon density is obtained from such an
inventory of observed baryons.

If we had a complete theory of how the baryons have been distributed
over various classes of astronomical objects at different epochs, as
structure in the universe developed, we
could in principle predict the cosmic density of baryons from the
measurement of only a single tracer of baryonic matter. Observations of
highly evolved virialized objects (galaxies and clusters) may not be
ideal candidates for such measurements, owing to the theoretical
uncertainties in the fraction of the total mass they contain, and in the
fate of the baryons which accreted onto these systems.  Here we shall
instead consider observations of gas in regions of much lower
densities outside virialized objects, which probably have a
more simple history.
Hydrodynamic simulations of increasing resolution allow us to
calculate the evolution of this cosmologically distributed gas from
initial perturbations down to redshifts accessible to observations, for
different cosmological models. 

The only way to observationally study the gas
at such low densities is by measuring resonance line absorption
imprinted on the spectrum of a background light source.
This phenomenon was first observed as the so-called Gunn-Peterson
effect (Gunn \& Peterson 1965): the increasingly redshifted \lya
absorption from intervening gas in the line-of-sight to a QSO causes an
apparent absorption trough blueward of the \lya emission line.  At high
spectral resolution the optical depth for \lya absorption is seen to
fluctuate sharply, giving rise to the observational phenomenon of the
\lya forest (Lynds 1971). The gaseous structures underlying the \lya
forest have often been visualized as discrete gas clouds producing the
absorption lines, embedded in a low density, distinct ``intercloud
medium'' that might cause a residual Gunn-Peterson absorption trough.
However, in a hierarchical structure formation picture where the \lya
forest originates in gravitational collapse, the photoionized gas
occupies a continuous, wide range of densities and pressures as it
accretes towards various structures, and there is no distinction
between a Gunn-Peterson effect and \lya forest absorption (Bi 1993;
Reisenegger \& Miralda-Escud\'e 1995; 
Hernquist \etal 1996; Miralda-Escud\'e \etal 1996; 
Croft \etal 1997; Bi \& Davidsen 1997).

  It is easy to see how the strength of the Gunn-Peterson effect must
depend on the baryon density and the photoionization rate. The optical
depth $\tau$ of HI \lya absorption in a quasar spectrum is proportional
to the neutral hydrogen column density and inversely proportional to
the velocity interval over which the gas is spread, $\tau \propto$
d$N_{HI}$/d$v$ (e.g.  Spitzer 1978). The neutral hydrogen density is
in turn proportional to $\Gamma^{-1}\, \alpha(T)\, n^2$, for highly
ionized gas dominated by photoionization.  Here $n_{HI}$ and $n$ are the
neutral hydrogen and the total gas density, $\alpha(T)$ is
the recombination coefficient, and
\begin{eqnarray}
\Gamma = 4\pi \int_{\nu_T}^{\infty} \frac{J(\nu)}{h\nu} \sigma_{\nu} d\nu \ \ s^{-1}
\end{eqnarray}
is the photoionization rate due to the background radiation with mean
intensity $J(\nu)$. A clumpy gas
with overdensity $\rho/\bar\rho$ and
temperature $T$, expanding with the Hubble flow, and  subject to peculiar
motions with velocity $v_{pec}(r)$ along the line-of-sight,
causes absorption with an optical depth

\begin{eqnarray} \tau \propto \frac{\left(\Omega_b H_0^2\right)^2}
{\Gamma \, H(z)}\, (1+z)^6 \, \alpha(T) \,
 \left(\frac{\rho}{\bar\rho}\right)^2
\left(1+\frac{dv_{pec}}{H_0 dr}\right)^{-1} ~, 
\label{eqn:tau}
\end{eqnarray} 
where $H_0$ is the present Hubble constant, and $H(z)$ is the Hubble
constant at redshift $z$ (ignoring here the effects of thermal
broadening). For a universe homogeneously filled with hydrogen gas
($\rho/\bar\rho$ = 1) expanding isotropically without peculiar motions
($v_{pec}$ = 0) we retrieve the original expression for the
Gunn-Peterson optical depth (Gunn \& Peterson 1965). But in a clumpy
universe, and because the observed flux decrement $D= e^{-\tau}$ is
extremely sensitive to the overdensity, the overdense regions with $\tau
\gtrsim 1$ can appear as distinct absorption lines, while most of the
spectrum between these lines contains only very weak absorption
features (corresponding to low-density gas filling most of the volume
in the universe), which are difficult to detect in the presence of noise.

Hydrodynamical simulations of structure formation 
can be used to generate simulated spectra as 
would be seen on a
source placed behind the simulated region of the universe, given the
peculiar velocities, temperatures, and densities in the absorbing gas
(Cen et al.\ 1994;
Zhang et al.\ 1995, 1996; Hernquist et al.\ 1996; Miralda-Escud\'e et al.\
1996, 1997). 
{}From a large number of simulated spectra we can calculate the
distribution of the optical depth, or the flux decrement $D=e^{-\tau}$. 
{}From equation (2), we see that once the gas overdensity, temperature and
velocity fields are fixed, only the normalization of the optical depth
$\tau$ in the spectra can be changed depending on the following
parameter:
\begin{eqnarray}
\mu \equiv \left({\Omega_b h^2 \over 0.0125}\right)^2
           \left({100\, {\rm kms}^{-1}\, {\rm Mpc}^{-1} \over H(z)}\right)
           \left({1 \over \Gamma_{-12}}\right) . \label{mu}
\end{eqnarray}
Here we have used $h = H_0/100 \kms\mpc^{-1} $, $\Gamma_{-12}=\Gamma/
(10^{-12} \sec )$, and 
the redshift dependent Hubble constant
\begin{eqnarray}
H(z) =  100 h\, {\rm kms}^{-1}\ {\rm Mpc}^{-1}\ 
\left[\Omega(1+z)^3 + (1-\Omega - \Lambda)(1+z)^2 + \Lambda\right]^{1/2},
\end{eqnarray}
where $\Omega$ is the total density in units of the critical density,
and $\Lambda$ is the contribution from the cosmological constant.
All cosmological parameters without explicit z dependence are given for 
z=0.

The parameter $\mu$ can be estimated by fitting the mean flux decrement
from the simulations to the mean flux decrement of the \lya forest
observed in QSO spectra.  Once this parameter is fixed, the shape of
the flux decrement distribution can be used to observationally test the model.

Note that the scaling of optical depths with $\mu$ assumes that
the gas temperatures, densities, and velocities do not change with
$\Omega_b h^2$, $H_0$, or $\Gamma_{-12}$, a point that we will return
to in our later discussion.
The optical depth is also
proportional to the term $[H(z)]^{-1}$ in equation (2), although a
change in the Hubble constant implies a change in the power spectrum of
the initial density fluctuations in the cosmological model as well.

  Thus, if we possess an independent estimate of the strength of the
ionizing radiation field, we can derive the value of $\Omega_b$
required to reproduce the observed mean \lya absorption for a given
cosmological model. If that cosmological model accurately reproduces
the other observed properties of the \lya forest, such as the column
density and Doppler parameter distributions and the distribution of
the flux decrement, then we can reasonably suppose that the effects of
the overdensities and peculiar velocities in equation (2) have been
adequately incorporated for the purpose of estimating the parameter
$\mu$, even if the cosmological model we use is not precisely the true
representation of the universe.

  The use of the \lya forest as a tracer of the baryons in the universe
has the advantage that the fraction of baryons present in the \lya
forest is predicted by the adopted cosmological model. Although there
are still uncertainties related to the fraction of baryons that may
have turned into stars at epochs earlier than the time when the \lya
forest is observed, this fraction is expected to be small (e.g.,
Couchman \& Rees 1986). It is reassuring that at redshifts $>$ 2 most
of the baryons are expected to reside in low and intermediate column
density condensations according to the simulations and that this is
consistent with the observations we have, so the \lya forest is not
only a tracer but the dominant reservoir of baryons (Rauch \& Haehnelt
1995; Miralda-Escud\'e et al.\ 1996, 1997).

  In this paper we compare the observed distribution of the flux
decrement $D$ from a new large dataset from the Keck telescope to the
predictions of two hydrodynamical simulations of the \lya forest. 
The first is the Eulerian simulation of a cold dark matter (CDM) model
with a cosmological constant
($\Lambda$CDM, with $\Omega$ = 0.4, $\Lambda = 0.6$, $H_0= 65
\kms\mpc^{-1}$, $\sigma_8=0.79$, and box size $L=10 h^{-1} \mpc$) analyzed by
Cen et al.\ (1994) and Miralda-Escud\'e et
al.\ (1996).  This model assumes a primordial power spectrum with an
asymptotic slope $n=0.95$ on large scales, and it is normalized to 
the COBE-DMR microwave background fluctuation amplitude.
The second simulation is the smoothed particle hydrodynamics (SPH) simulation
of the ``standard'' CDM model (SCDM, with $\Omega$ = 1, $\Lambda =0$,
$\sigma_8 = 0.7$, $H_0=50 \kms\mpc^{-1}$, and box size $L=11.1 h^{-1}
\mpc$) analyzed by Croft et al. (1997), which
is similar to the one presented in Hernquist et al. (1996) and Katz et
al. (1996) but includes star formation and a more realistic spectrum
for the ionizing background.  
This model has $n=1$ on large scales, but the $\sigma_8$ normalization,
chosen to yield a reasonable match to observed cluster masses, is only
about 60\% of the normalization implied by COBE-DMR for these cosmological
parameters.

We measure the parameter $\mu$ by scaling
the optical depth in the simulations so that the mean flux decrement
$\bar D$ agrees with the observed $\bar D$ at different redshifts.  We
then derive a lower limit to the ionization rate $\Gamma$ by requiring
that $\Gamma$ is at least as high as the number of ionizations caused
by the ionizing radiation field from the known QSOs.  {}From this a
lower limit to $\Omega_b h^{3/2}$ is obtained.  Furthermore, from the
redshift variation of the above parameter, we infer the redshift
dependence of $\Gamma$ which is consistent with each simulated
cosmological model.

  Section 2 describes the procedure we used to determine the flux
decrement distribution function from the observations. Section 3
presents the comparison between data and simulations, and the scaling
of the optical depth necessary to obtain agreement between them,
and Section 4
derives a lower limit for the intensity of the ionizing background.
In section 5 we discuss the consequences for $\Omega_b$ and for the evolution of the ionizing background, and in Section 6 we summarize our conclusions.

\section{The observed distribution of flux decrements}

The flux decrement distribution function (FDDF) was computed from a set
of 7 QSO spectra (described in Table 1) observed with the high
resolution spectrograph (HIRES) on the Keck
telescope. The nominal velocity resolution was 6.6 kms$^{-1}$ (FWHM),
and the data were rebinned onto $0.04\ {\rm \AA}$ pixels on a linear
wavelength scale.  The data were reduced as described by Barlow \&
Sargent (1997).  Continua were fitted to regions of the spectra judged
to be apparently free of absorption lines using spline functions. A
three-sigma rejection algorithm was used to eliminate statistically
significant depressions from the fitting regions.  The number of knots
used between the splines was dependent on the signal-to-noise and
redshift of the data. In the case of $z>4$ data only a relatively
crude low order polynomial with very few fitting points could be
used due to the lack of unabsorbed regions in the QSO spectrum
(with the corresponding large uncertainties in the absolute
continuum level).

We are interested in the distribution of optical depths for the Lyman
$\alpha$ $\lambda$1215.67 \AA\ absorption line, so only those regions of a
spectrum between the QSO's Ly$\beta$ and Ly$\alpha$ emission were
considered, to avoid confusion with the Ly$\beta$ forest. In addition,
spectral regions within 5 $h^{-1} \mpc$ from each QSO's emission
redshift were omitted to avoid contamination of the data sample by the
proximity effect (e.g., Bajtlik, Duncan, \& Ostriker 1988;
 Lu, Wolfe, \& Turnshek 1991).  The contribution
by metal lines to the opacity in the Ly$\alpha$ forest turned out to be
substantial in the low redshift ($z<$2.5) forest, amounting to
22\% of the total flux decrement at z$\sim 2$, so the spectra
had to be cleared of heavy element absorption lines and damped
Ly$\alpha$ lines.  Experience shows that the overwhelming majority 
\centerline{\bf Table 1: QSO Spectra Used}
\bigskip 
\centerline{
\begin{tabular}{cccr}   
QSO & $z_{em}$ & m \\\hline
Q2343+123& 2.52 & 17  \\
Q1442+293& 2.67 & 16.2  \\
Q1107+485 & 3.00 & 16.7  \\
Q1425+604 & 3.20 & 16.5  \\
Q1422+230 & 3.62 & 16.5  \\
Q0000---262 & 4.11 & 17.5  \\ 
Q2237---061 & 4.55 & 18.3  \\
\end{tabular}
}
of absorption lines with a ``narrow'' appearance (Doppler parameter less
than $\sim 15 \kms$) can be attributed to transitions of ions other
than HI.  Many of these lines can be identified  from other easily
recognizable lines at the same redshifts, redward of Ly$\alpha$
emission, but obviously there has to be residual contamination by systems
at lower $z$ (dominated by weak CIV systems) where most detectable
lines are buried in the Ly$\alpha$ forest.  Thus, whenever there was 
an unidentified strong and narrow ($b<10\kms$) line in the
forest it was cut out as well.
This approach is not totally satisfactory because weak metal lines
will still be left unidentified, but their contribution to the opacity
should be very small. On the other hand, there could also be some
narrow hydrogen lines that are mistakenly removed as metal systems,
but again we expect the contribution from any such lines to be
negligible.
With increasing redshift such residual errors decrease in importance
as the average opacity of the Ly$\alpha$ forest increases
rapidly, leading to increasing blanketing of any unidentified metal lines
while the strength (at least of the higher ionization) metal
lines decreases.

The spectral regions surviving the selection were split into redshift
bins ranging between $z$=1.5 and $z$=4.5, in 3 steps of $\Delta z$
=1.0.  To compare with the existing simulations the nominal central
redshifts were chosen to be 2.0, 3.0, and 4.0, but the actual mean
redshifts for the pixels used when computing the FDDF were $<z>$ =
2.29, 3.02, and 3.98, respectively. The discrepancy for the $z=2$ bin
results from a lack of low $z$ data in our sample, due to the low
sensitivity of the Keck HIRES instrument at wavelengths close to the
atmospheric cutoff.  At $\Delta z$ = 1, the redshift bins chosen are
rather wide and the mean redshift is obviously not always centered on
the middle of the bin, so evolution of the optical depth within each
bin may cause problems when comparing to spectra from a simulation at a
fixed redshift. The evolution of the optical depths from a freely
expanding  homogeneous medium in a flat universe should follow $\tau
\propto (1+z)^{4.5}$. Consequently, we have corrected the observed
optical depths within each bin by scaling them according to $\tau
\propto (1+z)^{4.5}$ to the values they would have at the central
redshift of the bin. The mean redshift for each bin, the original
observed mean flux decrement $\bar D$, and the corresponding value
after correction for the finite redshift range $\bar D^{cent}$ are
given in columns 2, 3, and 4 
of Table 2.

The FDDF was computed for each spectrum and each redshift bin
separately. Then the contributions from the 7 different spectra were
used to compute a weighted mean, where the median flux variance of each
spectrum $i$ divided by the number of pixels contributing to the
individual distribution, $\sigma^2_{med}(i)/N_{pix}(i)$, was used as a
weighting factor in the usual sense, i.e., the contribution to the FDDF
from a given spectrum was weighted in favor of spectral regions with
high average pixel signal-to-noise ratios and with many pixels in the
right wavelength range.  This procedure was necessary as the spectra
differed widely in wavelength coverage and in the average S/N ratio.

\begin{figure}[t]
\centerline{
\psfig{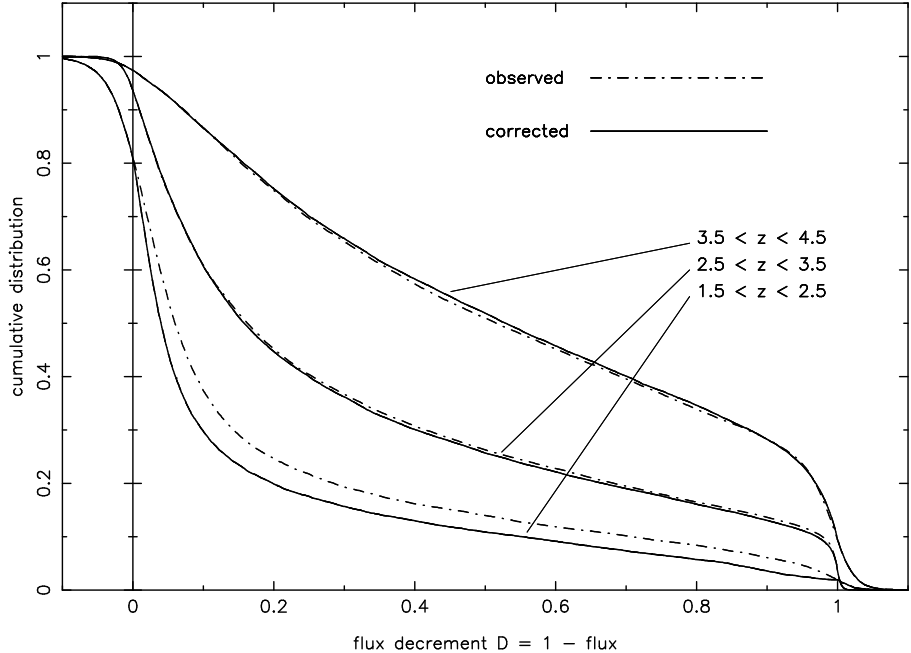}
}
\caption{\small The cumulative flux decrement distribution function (FDDF),
for the raw data (dash-dotted lines), and after a correction for the
evolution of the optical depth between the measured redshift and the
fiducial redshift (center of redshift bin) has been applied (solid
line).  \label{obscorr}} 
\end{figure}

  The resulting cumulative flux decrement distribution is shown in
Figure \ref{obscorr}. The FDDF for the raw data are represented by
dash-dotted lines, and the distributions for the corrected optical
depths are given by solid lines. Only the $z=2$ bin is changed
significantly by the evolution correction, 
because of the lack of low $z$ pixels in that bin (see
above).  The extent of the distributions below $D = 0$ and above $D = 1$
is due to the finite signal-to-noise ratio in the data, with the
bluer ($z=2$) curve having a flatter slope at these values because of the
higher noise level in the data.

\section{Comparison between observations and simulations}

\subsection{Corrections applied to the simulated spectra}

  In order to measure the parameter $\mu$ we choose to match a single
number, the mean flux decrement $\bar D$,  between observed data and
simulations, by scaling the optical depth by a constant factor. The
assumption we make here is that, had we repeated the simulations with
different values of $\Omega_b$ and $\Gamma$, the results would have
been identical except for a constant rescaling of the optical depth
at all points in the simulated spectra. 
Changing $\Omega_b$ and $\Gamma$ affects the cooling rates, photoionization
heating rates, and self-gravity of the gas, but most of the \lya
forest absorption arises in diffuse regions where the structure of the
gas distribution is determined by the gravitational potential of the 
underlying dark matter.  Changing $\Omega_b$ can systematically alter
the temperature of this diffuse gas by changing the photoionization
heating rate, but although this effect alters the index in the relation
$\tau \propto \Omega_b^\alpha$, the effect on the optical depth
distribution is still very close to a simple rescaling (see Croft
et al. 1997).  We also assume
that the neutral fraction in equilibrium
is proportional to the gas density, which is true if collisional
ionization is unimportant and the gas is highly photoionized. While
collisional processes are not negligible for high density regions,
these produce absorption with very high optical depth (i.e., strongly
saturated absorption lines) where changes in the optical depth do not
affect the flux decrement noticeably.
Thus the assumption of a linear scaling of the optical depth with $\mu$
(underlying eq.~[\ref{eqn:tau}]) 
should be a good approximation. We have used this
linear scaling on simulated spectra extracted from the two cosmological
simulations described in \S 1.  The pre-scaled spectra in the
$\Lambda$CDM model were generated assuming
$\Gamma=0.434\times 10^{-12} \sec^{-1}$, although the cosmological
simulation was run with a higher $\Gamma$ depending on redshift,
as described in Miralda-Escud\'e \etal (1996). 
The SCDM model was run with the photoionizing background computed
by Haardt \& Madau (1996) for $q_0=0.5$, but reduced in intensity
by a factor of two at each redshift (see Croft \etal 1997).

In estimating the parameter $\mu$ from the two simulations, an
additional problem arose from the fact that the fitting formulae for the
recombination coefficient that had been used in the analysis were
different in the Eulerian and the SPH simulation. When running the
numerical simulations, the accuracy of these formulae was not of great
concern because it did not significantly affect the physical evolution
of the gas. But of course, the neutral fraction that is calculated for
fixed physical density and temperature is proportional to the
recombination coefficient, and therefore the parameter $\mu$ that is
inferred is inversely proportional to the recombination coefficient that
is used. So, for the problem of deriving the $\mu$ parameter needed to
adjust the flux distribution from the simulations to the observed one,
it is important to use a recombination coefficient as accurate as
possible. The results of the most recent calculations for $\alpha(T)$
were given by Abel \etal (1996), and in the range
$3\times 10^3\, {\rm K} < T < 3 \times\, 10^4 {\rm K}$ (the relevant
range for the gas that contributes to the unsaturated regions of the
spectra), it is given by $\alpha(T) = 4.2\times 10^{-13} 
(T/10^4\,{\rm K})^{-0.7}$ cm$^3$s$^{-1}$ to within 3\%. The values of $\alpha(T)$ used for the
Eulerian (SPH) simulation differed from the formula of
Abel \etal (1996) by factors of 0.95 (1.20), which remain almost
constant over the same temperature interval. Therefore, to correct for
the slight inaccuracy in the recombination coefficients that had been used,
we simply divided the inferred values of $\mu$ by these factors for
the two simulations. More accurate expressions for the coefficients
determining the ionization will be incorporated in the future in the
numerical simulations, but the correction made here should be
sufficiently accurate for our purpose of determining $\mu$.
\begin{deluxetable}{cccccccc}   
\small
\tablewidth{0pt}
\tablenum{2}
\tablecaption{Observed and Corrected Flux Decrements}
\tablehead{
\colhead{$[z_1,z_2]$}& 
\colhead{$\bar z$} & 
\colhead{$\bar D$} & 
\colhead{$\bar D^{cent}$} &
\colhead{$\bar D^{cent}_{corr}$($\Lambda$CDM)} &
\colhead{$\bar D^{cent}_{corr}$(SCDM)} &
\colhead{$\bar D_{PRS}$} &
\colhead{$\bar D_{ZL}$} \\
\tablevspace{-.3cm}}
\startdata
\tablevspace{-.1cm}
1.5 - 2.5 & 2.29 & 0.186 & 0.148 &0.152&0.154& 0.15 & 0.08 \\
2.5 - 3.5 & 3.02 & 0.321 & 0.316 &0.330&0.345& 0.36 & 0.22  \\
3.5 - 4.5 & 3.98 & 0.539 & 0.543 &0.586&0.617& 0.62 & 0.63  
\enddata
\end{deluxetable}

We now describe the corrections that were applied to the observational
data in order to do the comparison of the flux distribution with the
simulated spectra.
Real data suffer from observational biases and errors that are not
present in idealized simulated spectra.
The effects of noise, instrumental resolution, and uncertainties
in the continuum level in the
observed data must either be taken out or imposed in a
similar way on the simulated spectra. In practice, ``degrading'' the
simulated spectra is usually easier.

The first and most important problem arises from our ignorance about the
precise placement of the QSO continuum against which the absorption
optical depth is to be measured. The usual manual fitting methods with
multiple splines or other high order polynomials tend to systematically
underestimate the zeroth or first order contribution to the continuum,
resulting in an underestimate of the number of pixels at low flux
decrement.  Moreover, at $z>3.5$ there appear to be very few pixels
left where the flux reaches up to the continuum within the noise
uncertainty (as is indeed expected to happen from the simulated
spectra), making the flux
distribution within 10-15\% of the true continuum even more uncertain.
Here we have adopted the very simple approximation of choosing
the highest flux value in each individual simulated spectrum (of length
determined by the size of the comoving periodic box of the cosmological
simulation) as the value of the
continuum, normalizing the spectra by dividing through this value.
Although there may be more accurate ways to correct for the
continuum fitting, the detailed process by which the continuum is
obtained is very complex to reproduce, and it would require us to obtain
longer simulated spectra than we have available from the simulation.
We have therefore adopted this
simple method because it should at least give an upper limit to the
effect of continuum fitting on the measurement of $\bar D$, since the
intervals that are fitted in the observed spectra tend to be 
longer than the simulated spectra.

To estimate the magnitude of the bias in $\bar D$ introduced when lowering
the continuum to the maximum flux level in each spectrum, we have also
computed the mean flux decrement for the models when retaining their
original continuum.  The difference between $\bar D$ derived for the
two different continuum settings can be used to predict a
correction for the actually observed $\bar D$, in the sense that the
``true'' observed $\bar D$ would have been larger if we had known
the position of the true continuum {\it a priori}. Thus, we define the
corrected, mean decrement $\bar D^{cent}_{corr}$ as the value of
$\bar D^{cent}$ in the idealized simulated spectra once the parameter $\mu$
is chosen to make the value of $\bar D^{cent}$ of the degraded, continuum
normalized spectra match the observed values. It represents our best
estimate of the true mean decrement assuming that the simulations 
give a realistic representation of the optical depth structure of the IGM.
The continuum bias corrections are larger for the SCDM model compared to
the $\Lambda$CDM model, because, as we shall soon show, the former model
has more low optical depth absorption.
The observed flux decrement corrected this way,
$\bar D^{cent}_{corr}$, is given in columns 5 and 6 of Table 2 for the
two models.
These values are in good agreement with the results obtained
by Press, Rybicki, \& Schneider (1993, PRS) with a different
method for a large sample of
low resolution QSO spectra, given for comparison in the last but one column 
of Table 2. However, the two lower z bins differ quite substantially
from the results of the analysis  by Zuo \& Lu (1993) 
(last column of Table 2), who find much less absorption at low redshift.   

\begin{figure}[t]
\centerline{
\psfig{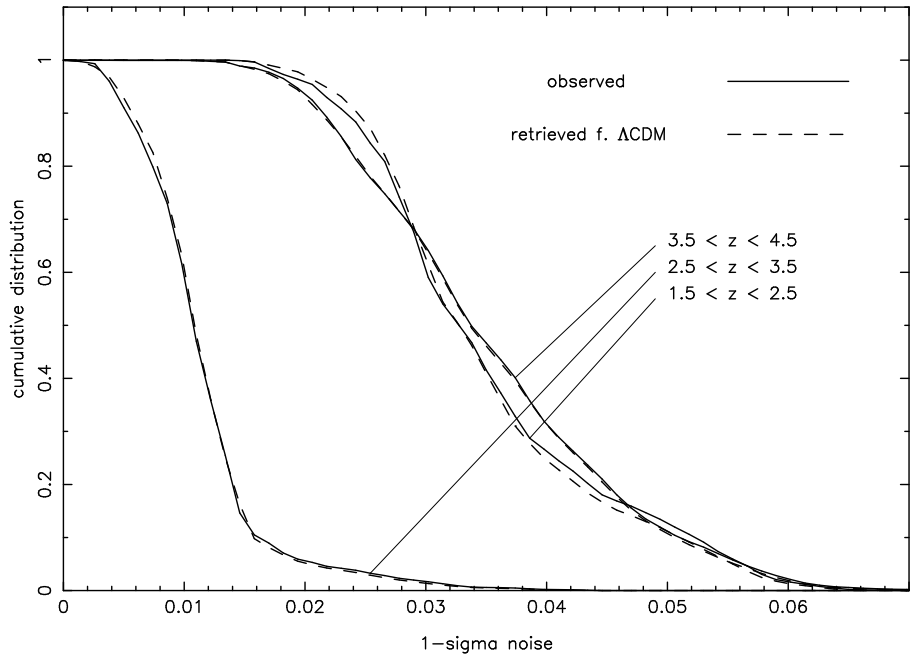}
}
\caption{\small Cumulative distribution of noise standard deviations for the
real data (solid line) and as derived from the simulated spectra
(dashed lines) of the $\Lambda$CDM model after applying the procedure
described in the text.\label{err}}
\end{figure}

The amount and distribution of noise is another important parameter,
critically  determining the shape of the FDDF close to the
continuum level (D$\approx$ 0) and zero level (D$\approx$ 1).
The total noise distribution of the real data is highly
variable and not necessarily close to any analytical shape.  There are
many reasons for this, the most important one being that the spectra
are a patchwork of individual exposures, taken under variable
conditions, convolved with a strongly varying sensitivity function of
the spectrograph.  Rather than attempting a detailed modeling of the
noise distributions and a decomposition into signal-dependent and
-independent noise, we have added noise to the simulations in the
following way: we take the noise variance arrays from the real data and
construct, for every observed spectrum and every flux level (in steps
of 5\%), the probability function for the noise variance. We treat these
functions statistically like the actual flux distribution functions for
the data (when forming weighted means), to obtain the resulting noise
probability function as a function of flux level. Then we add noise
consistent with the variances drawn from these probability functions to
the simulated data, using the appropriate distribution for each pixel
with a given flux level.  This gives an error distribution for the
simulated data very close to the real data, the only difference being
that the spatial correlation of the noise is of course lost, which is
irrelevant for the purpose of computing the flux distribution only.
Thus a pixel with flux decrement $D$ in the simulated data has the same
noise standard deviation as a pixel with the same flux in the real
data.  Fig. \ref{err} shows that the distribution of the noise
amplitudes retrieved from the simulations (here as an example from the
$\Lambda$CDM simulation) after applying this procedure (dashed lines)
is in excellent agreement with the distribution of the noise in the
real data (solid line). This is simply because the flux distribution of
the simulated spectra is very close to the observed one, as we shall
see later.

One additional correction has to
be made for the smoothing introduced when rebinning during the data
reduction.  The actual fluctuation in the data was typically between 10
and 40\% smaller than indicated by the error array produced from the
photon numbers right at the beginning of the data reduction before any
smoothing had occurred. Therefore, to make the simulated spectra look
the same as the real data, the noise fluctuations from the probability
distribution functions were reduced by suitable factors taken from a
comparison of the rms fluctuations and the error array in sample stretches
of the real
data. As a result the slopes of the FDDFs are well matched at both
ends, although  some discrepancy remains at the $D\sim$ 0 end because
of residual problems with the continuum level.

  The simulated spectra were also convolved with the instrumental profile,
which is a Gaussian with FWHM$= 6.6 \kms$, before the noise was added.
This convolution has a very small effect because all the features
appearing in the simulated spectra are well resolved at this
resolution.

\begin{figure}[t]
\vspace{-0.cm}
\centerline{
\psfig{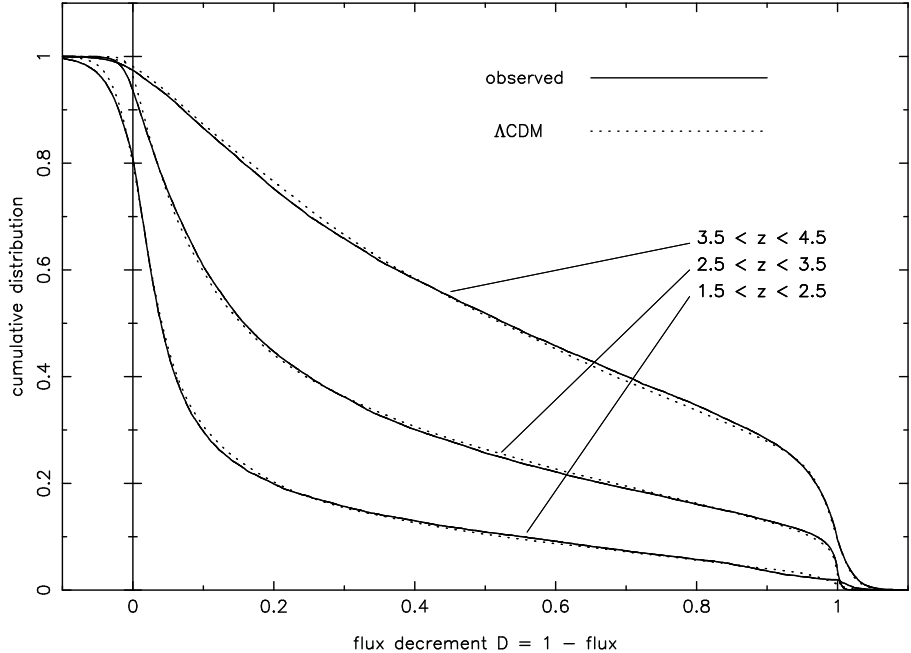}
}
\caption{\small The cumulative flux decrement distribution function (FDDF),
for the real data corrected for $\tau$ evolution as described in the
text (solid lines), and for simulated spectra (dotted lines) from the
$\Lambda$CDM simulation where the optical depth $\tau$  has been
scaled globally by the amounts given in Table 2. The simulated spectra
have had noise added and the continuum changed as described in the text.
\label{maxcont_99}}
\end{figure}
\vskip 1.cm
\pagebreak
\subsection{Results}

\begin{figure}[ptb]
\vspace{-0.cm}
\centerline{
\psfig{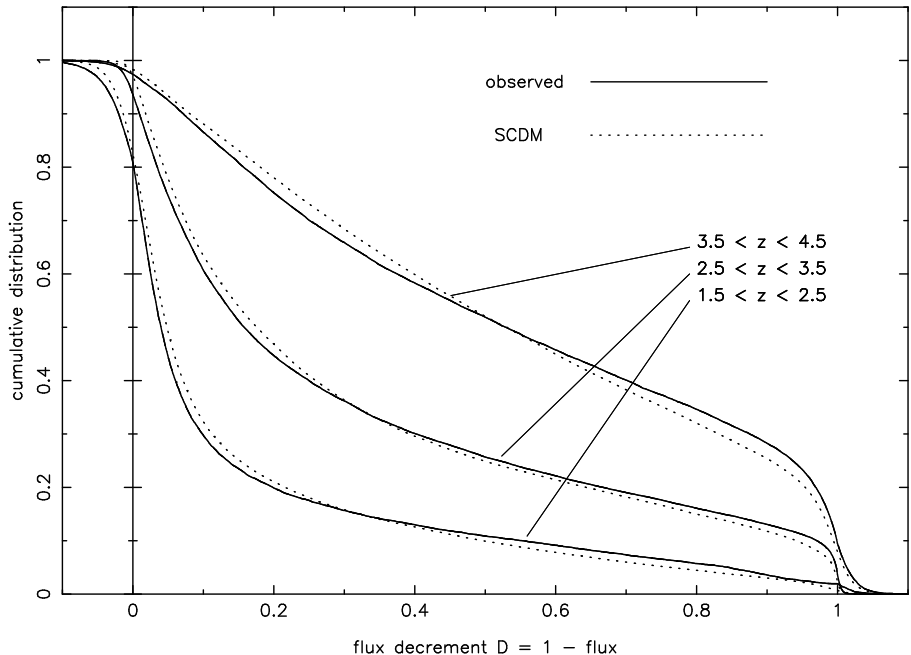}
}
\caption{\small The same diagram as before, but now for the
SCDM simulation. \label{maxcont_cdm_99}
}
\end{figure}

Figure \ref{maxcont_99} shows the FDDF for the $\Lambda$CDM
simulation (dotted line) overlaid on the observed distribution
(solid line). The results for the SCDM simulation are given in Fig.\
\ref{maxcont_cdm_99}, also compared with the observed distribution.
The agreement with observations is quite good in both cases,
given that it is the result of a one-parameter fit ($\mu$).
This agreement (together with the other characteristics of the
absorption lines that were found to agree reasonably with observations
as reported in previous papers) suggests that the theories of
hierarchical structure formation assumed in the cosmological
simulations provide us with an accurate physical picture of the \lya forest.
There are differences in detail in the predictions
from these two cosmological models.
\begin{figure}[t]
\centerline{
\psfig{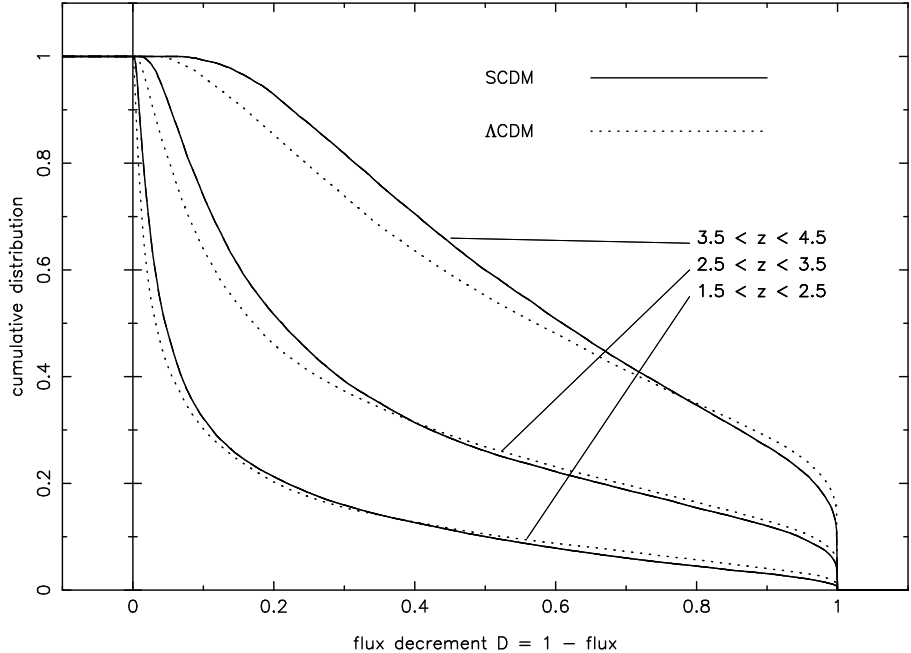}
}
\caption{\small Comparison between the shapes of the FDDFs for the $\Lambda$CDM (dotted
line) and SCDM (solid line) simulations, now for the raw spectra with the
original continuum level and no noise added.
The $\tau$ scaling is unchanged
from the previous two figures. \label{compare}
}
\end{figure}
In Figure \ref{compare}, the true FDDF of the SCDM and $\Lambda$CDM models are
plotted (i.e., before any corrections are made for the effects of
noise and continuum fitting), which allows us to compare the two
models directly without the alterations needed for comparison
to the observational data.  There is indeed a
small difference in the shape of the FDDF for the two models considered
here: the SCDM simulation has a larger contribution to the average flux
decrement from regions of low optical depth, compared to the
$\Lambda$CDM model.  Visual examination of Figures~\ref{maxcont_99}
and~\ref{maxcont_cdm_99} indicates that the $\Lambda$CDM simulation
fits the observations somewhat better than the SCDM simulation,
though assigning quantitative statistical significance to this 
difference will require a detailed examination of random and systematic
errors in the FDDF that is beyond the scope of this paper.

  What are the main characteristics of the gas distribution in the
numerical simulations that determine the resulting FDDF in the simulated
spectra? From equation (2), the optical depth distribution must be a
function of the distributions in overdensity and temperature of the gas,
as well as the effects that the peculiar velocity and thermal broadening
have in redistributing the optical depth from different spatial regions
in the observed spectra. The optical depth is highly sensitive to
the overdensity; since the overdensity is also distributed over a very
wide range, we expect that the distribution of overdensities
will be most important in determining the distribution of optical
depths. In fact, as shown in Miralda-Escud\'e \etal (1996, 1997),
the column density of absorption lines is mostly determined
by the overdensities of the intercepted structures. Given that the column
density distribution of the absorption lines agrees with the observed one,
it is then not too surprising that good agreement is found also in the
distribution of optical depths.

Therefore, one reason for the difference between the two models seen
in Figure \ref{compare} might be that the SCDM model has a higher fraction
of the gas in low density regions, giving rise to optical depths $\tau
\sim 0.2$, and a correspondingly lower fraction of gas in more overdense
regions that produce strong absorption lines ($\tau \gtrsim 0.7$).
The volume-weighted density distribution is plotted in Figure 6 for
the two models, at $z=2$. 

\begin{figure}[t]
\centerline{
\psfig{file=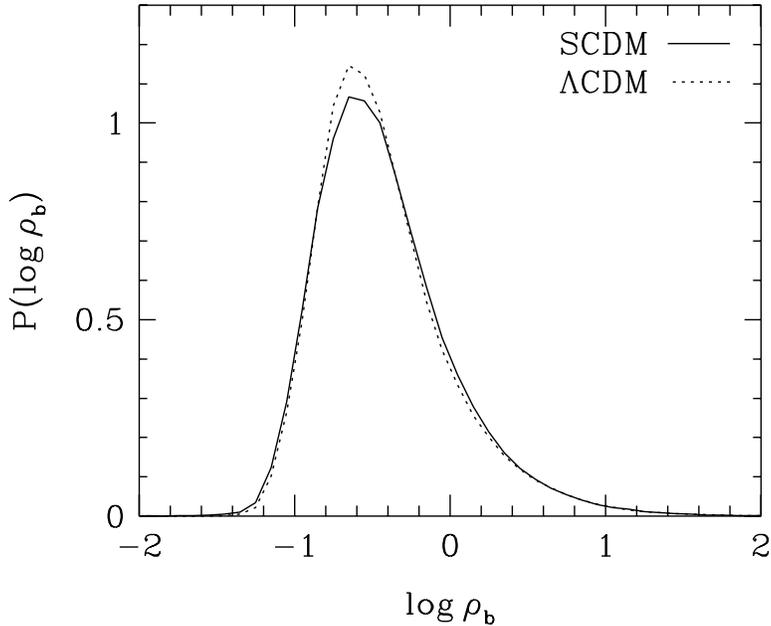,width=15.cm,angle=-0.}
}
\vskip -10.cm
\caption{\small The distribution of gas densities $\rho_b$ (in units of the
mean gas density) at $z=2$ for the SCDM simulation (solid line) and
the $\Lambda$CDM simulation (dotted line).\label{density}}
\end{figure}

The difference in the density distributions is very small, with the
$\Lambda$CDM model having a slightly lower dispersion in the overdensity.
This small difference has the wrong sign to explain the difference
in the FDDF: the smaller the dispersion in overdensities, the larger
the contribution to the mean decrement from unsaturated regions rather
than strong absorption lines.

  A second possible cause for the difference is the temperature of the
photoionized gas. The temperature affects the optical depth through the
value of the recombination coefficient, which is approximately
proportional to $T^{-0.7}$ in the range of interest. The median
temperature of the gas as a function of the overdensity is shown for
the two models at $z=2$ in Figure 7. At high overdensities, the temperature
is determined by shock-heating due to collapse of structure, and the
two simulations have similar temperatures. However, the temperatures
are different at low overdensities.
\begin{figure}[t]
\centerline{
\psfig{file=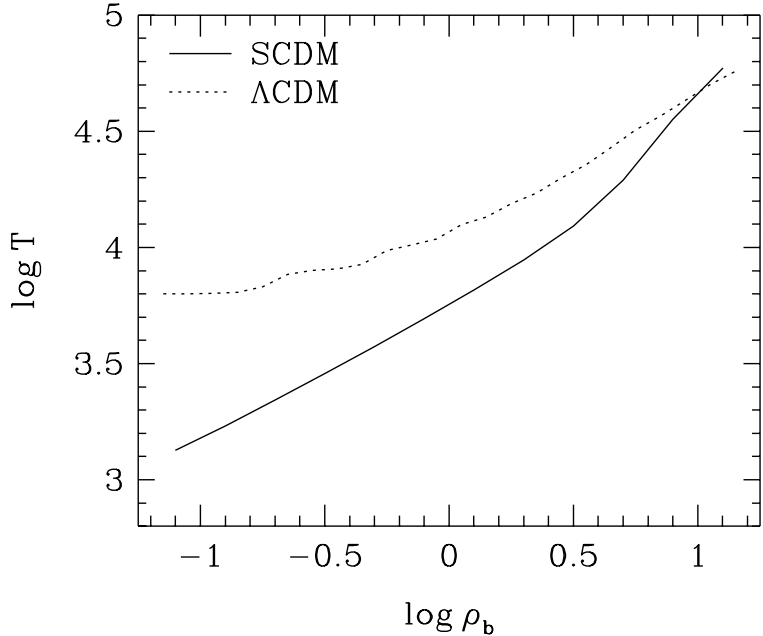,width=15.cm,angle=-0.}
}
\vskip -10.cm
\caption{\small The median temperature (in $^\circ$K) of gas with density $\rho_b$
(in units of the mean gas density) at $z=2$ for the SCDM simulation
(solid line) and the $\Lambda$CDM simulation (dotted line).}
\end{figure}

  The reason for this difference in temperature is that,
in the low density
regions, the gas temperature is not determined by photoionization
equilibrium alone, but it depends on the initial temperature which the
gas acquired when it was reionized (Miralda-Escud\'e \& Rees 1994).
Since the cooling time of the gas at low densities is longer than the
Hubble time, the gas retains a memory of these initial conditions. In the
SPH simulation of the SCDM model, the ionization is turned on at $z=6$
and the gas is assumed to be immediately ionized, but no heat is
included (the gas is only heated subsequently at the rate determined
assuming photoionization equilibrium). On the other hand, in the
Eulerian simulation of the $\Lambda$CDM model, initial heat from
the reionization is included; in fact, helium was doubly ionized at
$z\sim 3$ in this simulation (as a result of the relatively soft
spectrum that was assumed for the emitting sources), and this
resulted in an additional heating rate for low-density gas that was
absent in the SPH simulation. The result of these different assumptions
about reionization heating 
that were made in the two models 
is that in the Eulerian simulation the gas
temperature in the regions with $\rho/\bar\rho \lesssim 1$ is
significantly higher than in the SPH simulation, and
this difference in temperature increases as the density decreases.
Thus, the contribution to the average flux decrement from low density
regions in the SPH simulation is enhanced due to the higher
recombination coefficient relative to the Eulerian one.
Therefore, if the two simulations had
made the same assumptions on the heat deposited by reionization, the
changes in the distribution of optical depths in Figure~\ref{compare}
would improve the agreement.  We see that even the small difference 
found between the two models is probably not related to the
different cosmological models {\it per se}, but is explained in large part
by this difference in the gas temperature, which arises from the
differing treatments of reionization.

  The contribution of various physical effects to 
predictions for the FDDF will be examined in
greater detail in other papers (Weinberg \etal 1997).
Here, our main conclusion is that the two models we have examined 
appear to explain the observed flux distribution satisfactorily,
and therefore we
can use them to obtain the parameter $\mu$, which can provide us
with new constraints on the cosmic baryon density and the intensity
of the ionizing background.
\begin{figure}[t]
\centerline{
\psfig{file=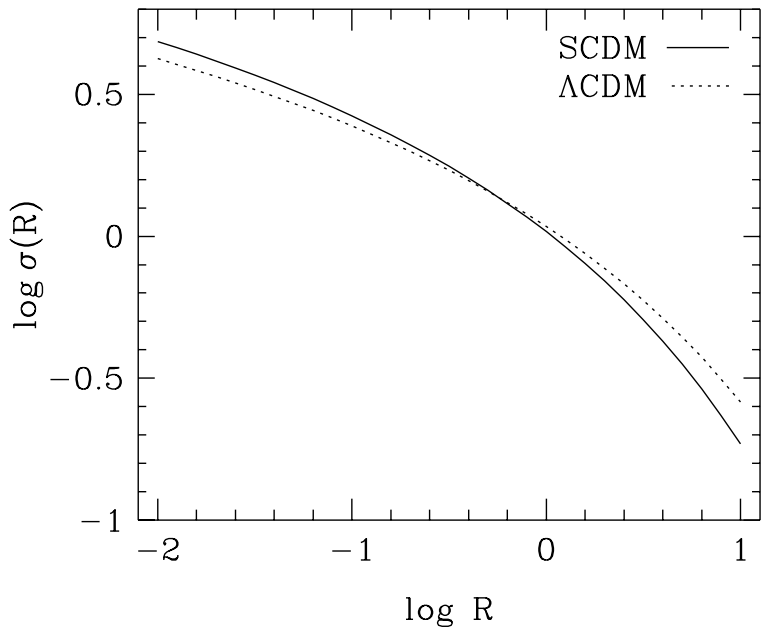,width=15.cm,angle=-0.}
}
\vskip -10.cm
\caption{\small The rms mass fluctuation in spheres of comoving radius 
$R\;h^{-1}\mpc$,
computed from the linear theory power spectra of the SCDM model
(solid line) and the $\Lambda$CDM model (dotted line) at $z=2$.
\label{fig:sigma}
}
\end{figure}

It is worth noting that the similarity in the FDDF predictions
of these two cosmological models is somewhat coincidental; they
predict similar nonlinear structure on the relevant scales at
$z \sim 2-4$, but some other popular theories of structure formation 
would not.  Figure~\ref{fig:sigma} shows the rms mass fluctuation 
as a function of comoving scale (tophat sphere radius $R\;h^{-1}\mpc$),
computed from the linear theory power spectra of the two models at $z=2$.  
The two models have quite similar mass fluctuation spectra on the
scales represented in these simulations, and the rms fluctuation 
amplitude is almost identical on the comoving scale $R \sim 1\;h^{-1}\mpc$
that is probably most relevant to the \lya forest at this redshift.
However, 
the rms fluctuation at $1\;h^{-1}\mpc$ 
for a COBE-normalized standard CDM model 
(with $\Omega=1$, $n=1$, $h=0.5$)
would be higher by about 70\%,
and it would be lower by about 30\% for a tilted CDM model
(with $n=0.8$, $\Omega=1$, $h=0.5$, $\sigma_8=0.55$)
and lower by more than a factor of two for a cold+hot dark matter
model with $\Omega_\nu=0.2$ (see figure~4 of Liddle et al.\ 1996).
These differences might well lead to significant departures from
the observed FDDF, though detailed analysis of these models 
will be required to see whether they are ruled out given the
freedom to adjust $\mu$ and the uncertainties in the appropriate
treatment of reionization (and in the resultant gas temperatures).

\section{
Observational Constraints on the Intensity of the Ionizing Background}

 We now proceed with examining independent constraints that we have on 
the intensity of the ionizing background at high redshift (or,
equivalently, the photoionization rate $\Gamma$), which we
can then combine with our determination of the parameter $\mu$ to
obtain limits on $\Omega_b$.

  There are several ways to measure observationally the intensity of the
ionizing background at high redshift. In this paper, we are
particularly interested in obtaining a firm lower limit to this
intensity, $J$. As we have argued above,
this leads to an interesting lower limit to the baryon
density $\Omega_b$. The best way to obtain a lower limit to $J$ is to
calculate the intensity from the observed number of sources and
absorbers of ionizing photons.  An important point here is that, in
principle, the observation of the number of sources and absorbers
determines the intensity of the background at all redshifts
independently of the cosmological model and any other assumptions. This
point becomes obvious by noticing that the background intensity would
not change if photons were emitted homogeneously in space, rather than
from individual sources; thus, only the average surface brightness from
sources in a given redshift interval matters for the calculation of the
background, and the surface brightness always varies as $(1+z)^4$.
Similarly, the absorption on the background depends only on the average
number of absorbers of a given optical depth per unit redshift along
the line of sight, which is also directly observed.

Let us express this directly using the equation for the evolution of
a cosmic background of proper intensity $J(\nu,z)\, d\nu$ at frequency
$\nu$ and redshift $z$ (e.g., Peebles 1971),
\begin{equation} (1+z)\, {\partial J(\nu,z)\over \partial z} =
- \left( \nu {\partial J\over \partial \nu} - 3J \right)  -
{c\over H(z)} \left( {j_E\over 4\pi} - \kappa J \right) ~.
\label{jev}\end{equation}
Here, $j_E(\nu)$ is the proper emissivity (energy emitted per unit time,
volume and frequency), $\kappa$ is the opacity, and $H(z)$ is the Hubble
constant at redshift $z$. If $\phi(L,z)$ is the comoving luminosity
function of sources (i.e., the number of sources per unit comoving
volume having a certain luminosity) at redshift z,
we have $j_E = (1+z)^3\, \int dL
\phi(L)\, L $. The observed flux per unit frequency from a given source
is $F(\nu) = L[\nu(1+z)] (1+z)/(4\pi D_l^2)$, where $D_l$ is the
luminosity distance. The observed total number of sources over the sky
at redshift $z$ and flux $F$ is related to the luminosity function by
\begin{equation}
N_s(z,F)\, dz\, dF = \phi(L)\, { 4\pi D_l^2\over (1+z)^2} \,
(1+z)\left( -{c\, dt\over dz}\right)\, dz\, dL ~.
\end{equation}
Finally, using $H= - dz/dt/(1+z)$, we obtain
\begin{equation}
J_E(\nu,z) \equiv {cj_E\over 4\pi H} = c(1+z)^4
\int_0^\infty dF\, {N_s\left( z, F[\nu/(1+z)] \right) \over 4\pi }\, F ~.
\end{equation}
The quantity $J_E$ is the one that is independent of the cosmological
model and is determined by the observations of $N_s$, and determines
also the evolution of $J$ through equation (\ref{jev}).

We can obtain the quantity $J_E$ from the results of various
quasar surveys that have been published. We use the results on
the quasar luminosity function of Warren, Hewett, \& Osmer (1994; their
Table 4), hereafter WHO, and
Hartwick \& Schade (1990; their Table 5), hereafter HS. We add up the
contribution to $j_E$ from each bin in luminosity of the luminosity
function given in these tables, and use eq.\ (2) in WHO, and the
relation $M_B = M_C - 0.605$ (see Pei 1995, for the model with spectral
index $\alpha = -0.5$ between the blue band
and \lya line wavelengths)
to transform the absolute magnitudes given in these papers to a
continuum flux per unit frequency at the \lya wavelength.
We then multiply this flux by a factor $(4/3)^{-1.5}$ to
transform to the flux per unit frequency at the Lyman limit, assuming
a spectral index $\alpha=1.5$ between the \lya and Lyman limit
frequencies (see, e.g., Laor \etal 1994). Table 3 shows
the values derived for $J_E(\nu)$ at the Lyman limit,
in the customary units of $10^{-21}$ \ergs ,
for two redshifts from WHO (we have
used their comoving luminosity functions in the intervals $2 < z < 2.2$,
and $2.2 < z < 3.0$, for calculating $J_E$ at $z=2$ and $z=3$,
respectively), and for $z=2$ from Hartwick \& Schade (where we use their
luminosity function in the interval $1.9 < z < 2.2$). The errorbars
for $J_E$ were obtained from the number of quasars from which the
luminosity function was derived in each luminosity bin. We also give
the result derived from the model of Pei (1995), which was used by
Haardt \& Madau (1996, hereafter HM)
to calculate the evolution of the ionizing background.

  We notice that $J_E$ derived from the luminosity functions of WHO
and HS agree with each other, but are lower by
a factor $2 - 3$ compared to the model in Pei (1995). The reason for
this is that the model by Pei was a fit to the observed luminosity
function which assumes $\phi(L) \propto L^{-1.83}$ at low luminosities
(for the model used in HM), and therefore the contribution
to the emissivity converges only as $L^{0.17}$. Since the value of the
luminosity below which $\phi(L)$ has this slope is not much higher than
the faintest quasars in the observed samples at $z>2$, a large fraction
of the emissivity in the model by Pei comes from quasars that have not
been observed, but are only assumed to exist in the model. 
One should
also notice that the luminosity function of HS extends to $0.6$
magnitudes fainter than in WHO. While the extrapolation of the
luminosity function in Pei (1995) to fainter quasars than observed is
reasonable, based on the luminosity function observed at lower
redshifts, we cannot rely on it to obtain a firm lower limit to $J_E$.
The values obtained by adding the contribution only from observed
quasars should be used as a lower limit, although the value from the
model of Pei should be considered as a more probable one. 
A considerably larger value of $J_E$ is always possible if sources of
ionizing radiation different from quasars are significant.

  HM derived the value of $J$, essentially using
equation~(\ref{jev}) but including absorbing clouds as sources in $j_E$ as
well, using the model of Pei for $J_E$. To obtain a firm lower
limit on $J$, we notice that if $J_E$ is decreased by a constant
factor at all redshifts $J$ will decrease by the same factor. This is
strictly correct as long as $\kappa$ is not altered, as seen from 
equation~(\ref{jev}). In the model of Pei, the
comoving density of quasars is assumed to decline at $z> 3.5$, so the
contribution from quasars above this redshift to the intensity $J$ at
$z=2$ is negligible (especially because of the increase of absorption
by Lyman limit systems at these redshifts). Thus, the value of $J$
cannot be significantly decreased by increasing the rate of decline of
quasars at high redshift, while being consistent with the observational
evidence that any decline of the quasar density is not significant below
$z\simeq 3.5$.
\begin{deluxetable}{ccc}
\small
\tablewidth{0pt}
\tablenum{3}
\tablecaption{Emissivities}
\tablehead{
\colhead{Ref.} & 
\colhead{Redshift} & 
\colhead{$J_E$} \\
\tablevspace{-.3cm}
}
\startdata
\tablevspace{-.1cm}
WHO & 2 & 1.50 $\pm$ 0.22 \cr
HS & 2 & 1.86 $\pm$ 0.69 \cr
Pei & 2 & 3.27 \cr
WHO & 3 & 2.92 $\pm$ 0.72 \cr
Pei & 3 & 7.09 \cr
\enddata
\end{deluxetable}

  The only other possibility there is for reducing the value of $J$ is
to increase the absorption $\kappa$. The calculation of HM included the
effect of absorption from a model of the number of absorbers given by
their equation~(7), which implies a number of Lyman limit systems per unit
redshift (with $\nhi > 1.59\times 10^{17}\cm^{-2}$) equal to
$1.5 [(1+z)/3]^{1.5}$. The observational determinations are consistent
with this number and are accurate to within $\sim 20\%$
(Sargent, Steidel, \& Boksenberg 1989; 
Storrie-Lombardi et al.\ 1994; Stengler-Larrea et al.\ 1995); since the absorption is dominated by
Lyman limit systems, it does not seem that absorption could be
significantly increased above the model used by HM. 
\dw{I remember that Piero told me that the main way to increase the
absorption would be to increase $f(N)$ in the range $10^{16}-10^{17}$,
since this is poorly constrained by observations and the systems
are strong enough to produce significant absorption.}
The reemission of
radiation from the absorbing clouds was also included in HM, and this
resulted in a substantial increase in the derived value of $J$.
The calculation of this reemission is also
model-independent and depends only on the observed column density
distribution of the absorbers, since the only assumption that is made
to infer the emission from clouds is that they are in ionization
equilibrium. This assumption is correct, because the time to reach 
ionization equilibrium depends only on $J$ and is of order $\sim
3\times 10^4$ yr. Thus, the inclusion of the reemission from clouds
to obtain a lower limit to $J$ is warranted.

HM obtained a value for the photoionization rate at $z=2$, $\Gamma =
1.4\times 10^{-12} \sec^{-1}$ (see their Figure 6).  From Table 3,
our lower limit to $J_E$ (taken from the average of WHO and HS) is a
factor of 2 below that in the Pei model used in HM, so we infer
\begin{equation}  \Gamma(z=2) > 7\times 10^{-13} \sec^{-1} ~.
\label{gam} \end{equation}
This can also be expressed in terms of the
cross-section-weighted background intensity (as defined in eq.\ [1] of
Miralda-Escud\'e \etal (1996)), $J_{HI} > 1.6 \times 10^{-22}$ \ergs . For the HM spectrum this
corresponds to an intensity $J_{912\AA\ }\approx 2.3\times10^{-22}$
at the Lyman limit.  

  One of the effects that can change the ionizing background intensity
from quasars is the possibility that they are obscured by
dust in intervening galaxies. However, it is simple to see that this
can only result in an {\it increase} of $J$ (see Miralda-Escud\'e \&
Ostriker 1990): the absorption from a quasar at $z=z_q$ to us must on
average be larger than from the same quasar to a point at $0 < z <
z_q$, so the increase in $J_E$ due to the fact that we underestimate
the number of quasars because they are obscured is more important than
the reduction of $J$ due to additional absorption. The observed fluxes
from quasars can also be altered by gravitational lensing, but this
does not affect the estimate of $J$ contributed by quasars because the
average surface brightness is conserved.

  We therefore conclude that our lower limit of eq.(\ref{gam}) is a strict
one, not subject to systematic uncertainties other than any errors in
the observational determination of the quasar luminosity function and
the number of absorbers.

  Another method for measuring the intensity $J$ is the proximity effect.
Observations at intermediate redshift
(Carswell et al.\ 1987; Bajtlik, Duncan \& Ostriker 1988; Lu, Wolfe
\& Turnshek 1991; Bechtold 1994; Giallongo et al.\ 1996; Cooke et al. 1996)
have arrived at values of $J_{912\AA\ }\sim 10^{-21}$, but with a
large scatter. The most recent high resolution studies by Giallongo et al.\
(1996) give $J_{912\AA\ }$= $(5\pm 1)\times 10^{-22}$ obtained for a redshift
range z=1.7-4.1, and Cooke et al. find $(10^{+0.5}_{-0.3})\times 10^{-22}$
for a similar $z$ range. However, the proximity effect is subject to several
systematic uncertainties; in particular, $J$ could be underestimated
if the luminosity of quasars is highly variable over the ionization
timescale of $3\times 10^4$ yr, and also if most of the quasars with
good spectra (which are naturally the brightest ones) are magnified by
gravitational lensing. In addition, the number of clouds near quasars
might be enhanced due to clustering, partially cancelling the reduction
due to the higher intensity of photons. It is difficult to estimate how
large these effects could be, so the proximity effect cannot be used to
obtain a firm lower limit to $J$. Nevertheless, the fact that all the
estimates are higher by factors of $2$ to $20$ compared to our lower
limit from the number of observed quasars is reassuring.

An independent, rough estimate of $J_{912\AA\ }$ can also be derived
from the degree of ionization needed to reproduce the column density
ratios of various metal ions in intervening heavy element absorption
systems. SCDM simulations specifically adressing  such higher column
density systems (Rauch, Haehnelt, \& Steinmetz 1996) have shown that at
z $\sim$3 a HM spectrum with an intensity $J_{912\AA\ }$ =
$3\times10^{-22}$ matches well the observed ratios  of several common
metal ions (for a uniformly metal enriched gas with Z=$10^{-2.5}\odot$,
$\Omega_b h^{2}$ = 0.018 ).  Hellsten \etal (1997) have confirmed this
result for the SCDM simulation described here.  Again, this is fully
consistent with our lower limit.

\section{ Discussion}

\subsection {The lower limit to the cosmic baryon density}

  With the lower limit on $\Gamma$ we immediately obtain a lower limit
for $\Omega_b$, using the constraints from the simulations. Our lower
limit for $\Gamma$ is applicable at redshifts $z=2$ and 3, since this is
the redshift range where the sources have been observed and the models
are consistent with an approximately constant intensity contributed by
the known sources (see HM). The constraints for any changes in $\Gamma$
when going to $z=4$ are relatively poor, because of the small number of
known quasars at this high redshift. In Table 4, we give the values of
the parameter $\mu$ (eq.~[\ref{mu}]) derived from our fits to the FDDF in
Section 3 at each redshift and for the two models. The limits on
$\Omega_b h^2$ assuming $\Gamma > 7\times 10^{-13} \sec^{-1}$, as
required from the two simulations, are given in the fourth and fifth column.

\begin{figure}[t]
\centerline{
\psfig{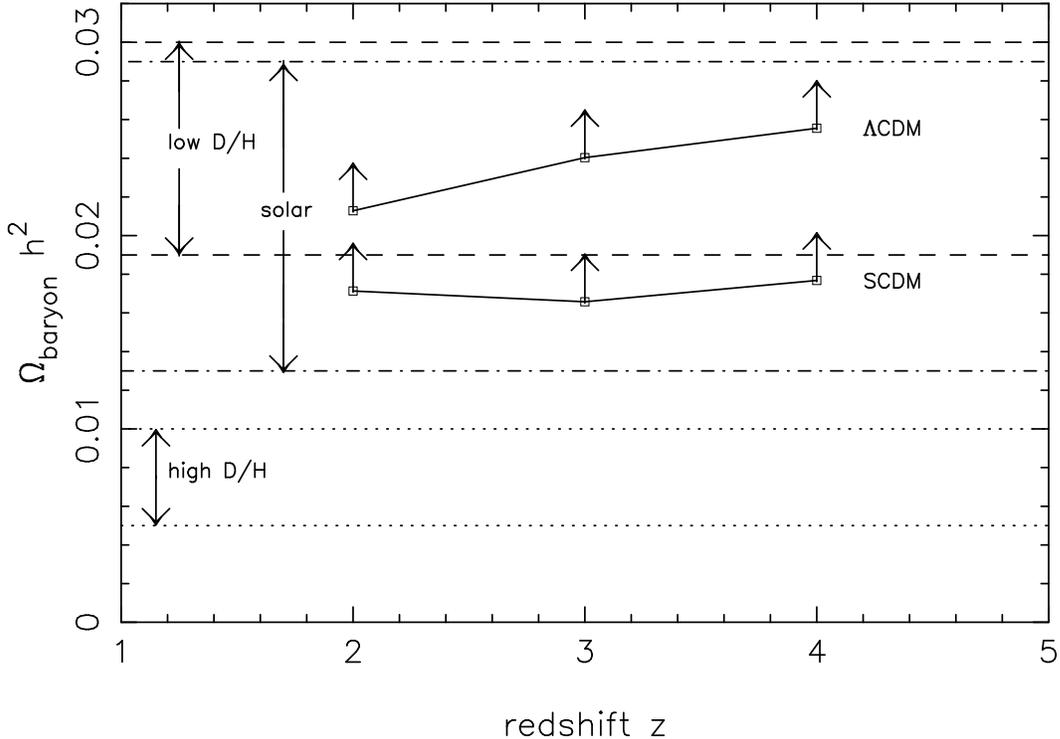}
}
\caption{\small Lower limits to $\Omega_b h^{2}$ assuming $\Gamma >$
7$\times10^{-13}$ s$^{-1}$. The region between the dashed lines is the
95\% confidence area for the ``low D/H'' value as derived by Hata et
al. (1996). The dotted lines show the corresponding lower and upper
limits for the ``high D/H'' value, and the
dash-dotted lines give the value consistent
with solar system D and $^3$He abundances, according to the same source.  
Note that the nucleosynthesis constraints measure
$\Omega_b h^{2}$, whereas our method gives
$\Omega_b h^{3/2}$.  \label{omega}}
\end{figure}
\begin{deluxetable}{cccccccccc}
\small
\tablewidth{0pt}
\tablenum{4}
\tablecaption{Limits on $\Omega_b$ and $\Gamma$}
\tablehead{
\colhead{} &
\colhead{} &
\multicolumn{2}{c}{$\mu$} &
\colhead{} &
\multicolumn{2}{c}{$\Omega_b h^{2}$ (for $\Gamma>7\times10^{-13}$)} &
\colhead{} &
\multicolumn{2}{c}{$\Gamma_{-12}$ (for $\Omega_b h^{2}$ = 0.024)} \\
\tablevspace{.2cm}
\cline{3-4}\cline{6-7}\cline{9-10}\\
\tablevspace{-.3cm}
\colhead{$[z_1,z_2]$} &
\colhead{} &
\colhead{$\Lambda$CDM} & 
\colhead{SCDM} &
\colhead{} &
\colhead{$\Lambda$CDM} &
\colhead{SCDM} &
\colhead{} &
\colhead{$\Lambda$CDM} & 
\colhead{SCDM}\\
\tablevspace{-.4cm}
}
\startdata
\tablevspace{-.1cm}
1.5 - 2.5 & &1.888 &1.033 & &$>$ 0.0213&$>$ 0.0171&&0.890&1.373\\
2.5 - 3.5 & &1.587 &0.628 & &$>$ 0.0240&$>$ 0.0166&&0.698&1.468\\
3.5 - 4.5 & &1.291 &0.511 & &$>$ 0.0255&$>$ 0.0177&&0.618&1.290\\
\enddata
\end{deluxetable}
In Figure \ref{omega}, we plot these same limits together with the 95\%
confidence limits from Hata et al.\ (1996) for the $\Omega_b$ derived
for the cases of a ``low''  and a ``high'' D/H ratio recently discussed
in the literature, and for a standard big bang model with solar system
D and ${}^3$He abundances. On the basis of profile fits to DI and HI
\lya absorption lines of intermediate column density systems at high
redshift, Songaila et al. (1994), Carswell et al.  (1994), and Rugers
\& Hogan (1996a,b) reported a high D/H ratio, whereas Tytler, Fan, \&
Burles (1996) and Burles \& Tytler (1996) favored a low D/H ratio as
representative of the universal deuterium abundance.
 If the simulations are correct, the low $\Omega_b$ value
(corresponding to the high D/H) is not consistent with the higher
values we obtain here for the two CDM models.  The range for solar
system abundances, and the lower D/H value (Tytler, Fan \& Burles 1996)
are fully consistent, as our measurements are lower limits to
$\Omega_b$.

  The obvious question that should be asked here is if other cosmological
models of the \lya forest might require a much lower value for the
parameter $\mu$ to fit the observed FDDF, while still predicting
satisfactorily the observed characteristics of the \lya forest.
As discussed in Miralda-Escud\'e \etal (1996), in order for that to be
possible, the physical structures giving rise to the \lya absorption
systems of a fixed, observed neutral hydrogen column density would need
to be much denser than they are in the simulations we have examined here,
while still producing lines of the same velocity width (and hence
equivalent width).
To clarify this argument, we shall give an analogous one here that will
focus on the distribution of the flux decrement rather than
the number of absorption lines with fixed column density.

  Let $f(\tau)\, d\tau$ be the optical depth probability density, i.e.,
the fraction of the \lya forest spectrum where the optical depth is in
the range $\tau$ to $\tau + d\tau$. Since $f(\tau)$ is related to the
FDDF in a straightforward way, we know that our simulations reproduce
reasonably the form of $f(\tau)$. Now, consider the spatial regions
in the simulation that yield optical depth $\tau$ to $\tau + d\tau$
in the \lya spectrum, and let $f_r(\tau)\, d\tau$ be the fraction of
the spatial volume filled by these regions.
The average optical depth contributed by such regions is:
\begin{equation} \tau\, f(\tau)\, d\tau \propto
\mu (\rho/\bar\rho)^2\, \alpha(T)\, f_r(\tau)\, d\tau \propto 
\mu (\rho/\bar\rho)\, \alpha(T)\, f_b(\tau)\, d\tau ~,
\label{taucon}\end{equation}
where $\rho/\bar\rho$ and $\alpha(T)$ are the overdensity and the
recombination coefficient in the regions that give optical depth $\tau$,
and $f_b(\tau) = f_r(\tau)(\rho/\bar\rho)$ is the fraction of
baryons in these regions. One should notice that, in general, the
optical depth at a given point in the \lya spectrum will not come from
a unique point in space, but from a finite region, owing to the effects
of thermal broadening and velocity caustics; at the same time, regions
yielding certain optical depths will have a distribution of densities
and temperatures. But for the purpose of the present argument, the
density and temperature in equation (\ref{taucon}) should be understood as
a representative value for the gas that contributes to optical depths
$\tau$.

  We see from equation (\ref{taucon}) that, if $\mu$ were to be lower
than the value we found from our two simulations (to allow for a
smaller $\Omega_b$), then either a larger fraction of baryons ought to
be in the intergalactic gas producing the \lya forest, or the
overdensities of the structures should be higher, or the temperatures
should be lower to increase the recombination coefficient. The first
possibility cannot make a large difference, because most of the baryons
in our simulations are already in the \lya forest (Miralda-Escud\'e et
al.\ 1996, 1997). The inferred $\mu$ also cannot be significantly
decreased by having a lower temperature of the gas, because in the SPH
simulation the gas temperature in low density regions is already as low
as possible. No energy input from reionization was included in this
simulation, and the temperature is then determined by the balance
between adiabatic cooling and the heating from photoionization (with
other cooling terms being less important), and it does not depend on
the model for structure formation. This minimum temperature is
approximately proportional to $[\Omega_b h(z)]^{0.6}$, as will be
explained in more detail in Miralda-Escud\'e et al.\ (1997). 
The increase of temperature with $\Omega_b$ tends to strengthen our
lower limit to $\Omega_b$, since we have assumed that 
$\tau \propto \Omega_b^2$ while, in the case where reionization heat
input is small as in the SCDM simulation, the effective dependence
is shallower (Croft et al. 1997; Miralda-Escud\'e et al. 1997).

  The difference in temperature is in fact the primary reason that the
SCDM model yields a lower
value of the parameter $\mu$ compared to the $\Lambda$CDM model: most
of the increase of the mean decrement as the parameter $\mu$ is
increased comes from unsaturated absorption arising from regions with
$0.3 < \rho/\bar\rho < 1$, where the difference in temperature is a
factor $\sim 3$, implying a difference in the recombination coefficient
of a factor 2, similar to the difference between the $\mu$ parameters
in the two models from Table 4.

  This leaves only the possibility that the structures yielding the
observed \lya forest absorption are more overdense than predicted in
our simulations. That is to say, if we fix the optical depth $\tau$,
the absorption that fills a fraction of the spectrum $f(\tau) d\tau$
determined observationally must arise from gas with higher densities
than in the simulation, and therefore also with smaller filling factors
$f_r(\tau)$ in real space. We
notice that, since $\mu \propto \Omega_b^2$, and the absorption is
proportional to $\mu (\rho/\bar\rho)$ when $f_b(\tau)$ is fixed (from
eq.~[\ref{taucon}), then in order to decrease our
lower limit to $\Omega_b$ by a factor of 2, the
overdensities of \lya absorbers with a fixed baryon content should be
larger by a factor of 4 compared to our model. Since the shape of the
FDDF needs to be preserved, this increase in density would have to be
rather homogeneous for all the gas giving rise to different optical
depths. It is difficult to see how this could be achieved unless we
were to return to a picture of separate clouds, and an
``intercloud medium'' which does not make a significant contribution
to the observed absorption.

  Given this argument, we believe that the lower limit we obtain
here on $\Omega_b$ is unlikely to be much weaker in other cosmological
models, at least if they remain consistent with the observed FDDF.
Indeed, we have found that the difference in the values of $\mu$
inferred from the two models reflects
the difference in the temperature of the low-density gas in the two
simulations, which is a consequence of the treatment of reionization
rather than a consequence of the cosmological models {\it per se}.
If, as argued above, the temperature of the low-density gas cannot be
lower than in the SPH simulation, then our lower limit is
$\Omega_b h^2> 0.017$, from Table 4 (we do not consider the $z=4$ lower
limit reliable because of the more uncertain value of $\Gamma$).
If we accept that $\mu$ cannot be lowered below the value in this model,
we can still reduce the inferred $\Omega_b$ in a model where $H(z)$ is
as low as possible. If we take $H_0> 50 \kms\mpc$, and consider the
$\Lambda$ model to minimize the increase of $H(z)$ with redshift, with
$\Omega > 0.3$ to satisfy constraints from large-scale structure
(e.g., Dekel \& Rees 1994),
we can reduce $H(z=2)$ by a factor $0.57$ relative to the SCDM model,
which would bring down our lower limit to $\Omega_b h^2 > 0.012$. 
Even with all parameters pushed to their limits, 
this is still inconsistent with the high deuterium measurements.

\jm{ I have summarized the paragraph below from what was written previously
by David, he will need to check this when he gets back next week}

  Croft et al.\ (1997) have recently studied HI and
HeII \lya absorption in SPH simulations of several cosmological models,
including the SCDM simulation analyzed here,
an open CDM model with $\Omega_0=0.4$, 
and a COBE-normalized, $\Omega=1$ CDM model with $\sigma_8=1.2$
(instead of the $\sigma_8=0.7$ as adopted for the SCDM model).
The open model has a similar amplitude of fluctuations on the \lya
forest scales at $z \sim 2-3$ ($\sim 1h^{-1}$ Mpc), so not surprisingly
the results for the parameter $\mu$ and for the flux distribution
function are similar to the SCDM model. On the other hand, the
CDM model with $\sigma_8 = 1.2$ has a higher amplitude of fluctuations,
leading to a wider density distribution: more of the gas in this
model has collapsed into high density regions producing saturated 
absorption, and less gas remains in the low and intermediate density
regions giving the unsaturated absorption.
As a result, this model requires a higher value of $\mu$ to match
the observations, i.e., a larger $\Omega_b$.
The predicted FDDF is also correspondingly broader
(see Croft \etal, figure 11), perhaps at a level
that could be ruled out by comparison to the FDDF measured here.
In order to weaken our lower bound on $\Omega_b$, one would want a
cosmological model with a lower mass fluctuation amplitude than the
SCDM or $\Lambda$CDM models, as these would leave a larger fraction of
their gas in unsaturated regions. Such models certainly exist
(tilted CDM or cold$+$hot dark matter, for example), but it does not
seem likely that the fluctuations in the IGM can be reduced greatly
without spoiling agreement with the FDDF. These models would
also risk conflict with the number of observed damped \lya systems
(Katz \etal 1996; Gardner \etal 1997). The exact constraints given 
on the models by
the observed FDDF, the damped \lya systems, and other observables of
the \lya forest such as the Doppler parameters, together with
the allowed variations on the minimum value of the parameter $\mu$
required to reproduce the observed mean decrement, will be investigated
in more detail in future work.

\subsection{ The evolution of the ionizing background}

\begin{figure}[t]
\centerline{
\psfig{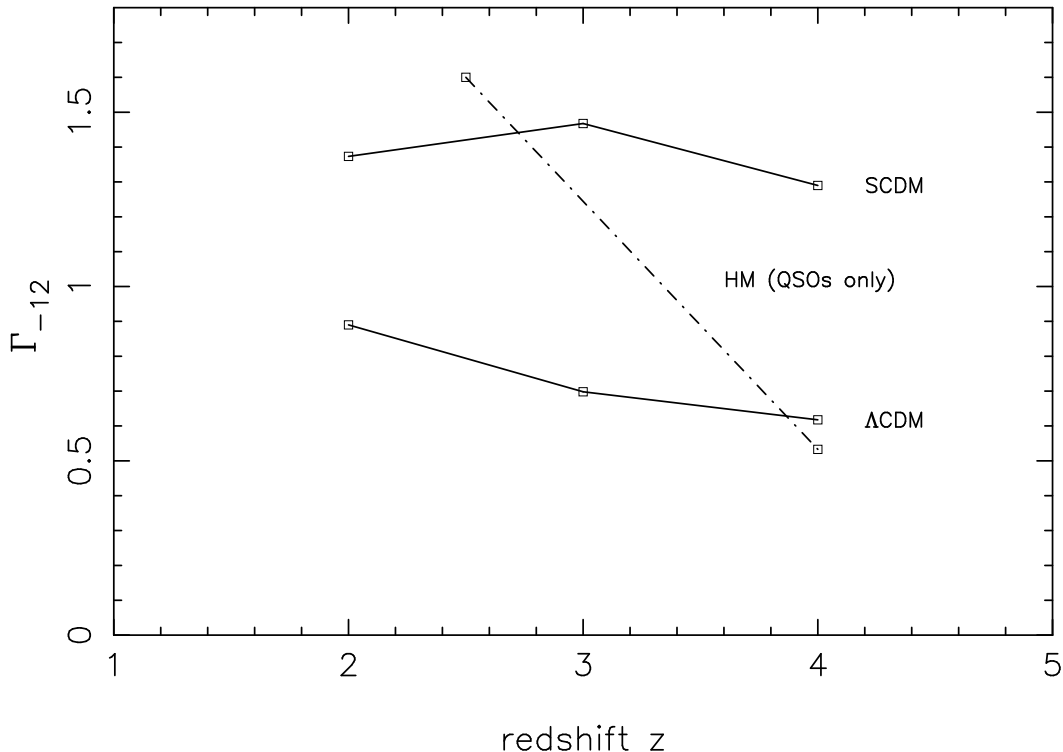}
}
\caption{\small Redshift dependence of the ionization rate $\Gamma$, adopting
a constant $\Omega_b h^{2}$ =0.024 (the value given by Burles \& Tytler 1996).
The dash-dotted line connects two estimates for $\Gamma$ by
Haardt \& Madau (1996), based on the expected contribution from QSOs alone.
The steep drop of their estimate towards z=4 and the near constancy
of the ionization rate required by our measurement indicate that additional
sources of ionization other than the known QSOs dominate beyond z$\sim$ 4.
\label{flux}}
\end{figure}

  The variation of the parameter $\mu$ with redshift can be used to
determine the evolution required for the intensity of the ionizing
background, or the photoionization rate $\Gamma(z)$.  For this purpose,
we choose $\Omega_b h^2 = 0.024$ to fix the normalization of
$\Gamma(z)$, which is consistent with the deuterium measurements by
Tytler \etal (1996).  The inferred photoionization rates are shown for
the two models in Figure~\ref{flux}. The intensity is almost constant,
with a slight decline with redshift required  for the $\Lambda$CDM
model.  This finding is consistent with measurements from the proximity
effect, where there is observational evidence for a small decrease in
the intensity above redshift 4 (Williger et al. 1994; Lu et al. 1996).
The values of $\Gamma_{-12}$ as a function of redshift are listed in
Table~4.

\vfill\newpage

Comparing to the model in HM for the intensity derived from the
observed quasar luminosity function and number of absorbers, we see
that the intensity is in good agreement at $z=2$ and $z=3$. It is
therefore remarkable that if the deuterium measurement of Tytler \etal
(1996) is correct, then the simulations predict the same value of
$\Gamma$ as expected from the observed sources of ionizing photons.
Since $\Omega_b$ cannot be substantially larger than the value assumed
here if primordial nucleosynthesis is correct (because even allowing for
a large systematic error in the ${}^4$He abundance, the solar system
deuterium abundance would then be in conflict with the theory),
our simulations then imply that
sources other than the observed quasars cannot contribute significantly
to $\Gamma(z)$, at redshifts up to three. On the other hand, the
decline of $\Gamma$ at $z=4$ in the HM model is not indicated by our
measurement.  Therefore, our models predict that sources of ionizing
photons other than just the known QSOs should be present at
$z\gtrsim 4$ to account for the cosmic background. The additional
emission of photons might come from other quasars that have not yet been
identified, or from star-forming galaxies.
An alternative possibility is that, for $z \ga 4$,
absorption by dust in intervening galaxies reduces the flux
from quasars seen at $z=0$ (used in the HM computation of $\Gamma$) 
relative to the flux seen by the IGM at $z \sim 4$
(Heisler \& Ostriker 1988; Miralda-Escud\'e \& Ostriker 1990).

\section{Conclusions}

Hydrodynamical simulations of hierarchical structure formation at high
redshift have now reached a degree of realism enabling us to determine
cosmological parameters from a comparison of the simulated HI
distribution with observational data about the \lya forest.  Here we
have compared the distribution of flux decrements in simulations and
observations, and we have measured the quantity $(\Omega_b h^2)^2/
[\Gamma H(z)]$ by scaling the optical depth distribution of the
simulated \lya forest spectra such as to match the mean flux decrement
$\bar D$ in the observed data.  As a first result we find that, in spite
of some individual differences, both a $\Lambda$CDM model (Cen et al.\
1994; Miralda-Escud\'e et al.\ 1996) and a standard CDM model (Hernquist
et al.\ 1996) are able to reproduce the basic shape of the cumulative
flux decrement distribution function well. Although not a proof, this
gives additional support to hierarchical structure formation as the
process that leads to the distribution of the baryonic density and
temperature in the universe as manifested in QSO absorption spectra.
This result is robust to the uncertainty in the allowed cosmological
model within the range tested here, though other cosmological models 
that predict substantially different high-redshift structure may
eventually be ruled out by comparison to the observed FDDF.

To measure $\Omega_b$ separately we have estimated  the ionizing
background radiation from the UV intensity produced by the known QSOs
alone.  The inferred ionization rate of neutral hydrogen,
$\Gamma > 7\times10^{-13}$ s$^{-1}$,
is a strict lower limit and is consistent with the lower range of
the intensities determined from the proximity effect in the \lya forest
near QSOs.  The lower limit on $\Gamma$ then translates into a lower
limit on $\Omega_b h^{3/2}$. The limits obtained from the two simulations
are different by a factor $\sim 1.5$, but we have shown that most of
this difference is related to the different temperature of the
low-density gas in the simulations. After considering the
uncertainty in this temperature, as well as the uncertainty in the
Hubble constant, we arrive at a lower limit $\Omega_b h^2 > 0.012$. As
we have discussed, this lower limit might be reduced in models with a
lower amplitude of density fluctuations, but it is doubtful that such
models would also agree with the observed distribution of the flux
decrement, and that they would be able to reproduce the observed mass
of baryons in the damped \lya systems.

  When we take the best estimate from the models of HM for $\Gamma$
contributed by known QSOs, $\Gamma\simeq 1.4\times 10^{-12}$ s$^{-1}$
(rather than the above lower limit), our inferred value of $\Omega_b$
for the two models we have examined is consistent with the
value implied by the deuterium abundance measurement by Tytler \etal
(1996), whereas our lower limit to $\Omega_b$ is inconsistent with the
much higher deuterium values found by others (Rugers \& Hogan 1996a,b
and references therein). We also conclude from this result that any
other sources of UV photons at high redshift could not increase the
value of $\Gamma$ to more than $\sim 3\times 10^{-12}\, {\rm s}^{-1}$,
because the very large $\Omega_b$ that would then be implied would be
inconsistent with primordial nucleosynthesis, given the Solar System
deuterium abundance.

  From the redshift dependence of the parameter $\mu$ in our two models,
we can also infer the required evolution of $\Gamma$. We derive an
approximately constant value from $z=2$ to $z=4$, which is
in agreement with the recent measurements of the proximity effect.
The decrease of $\Gamma$ between $z=3$ and $z=4$ expected if the
ionizing background were produced solely by the known population of QSOs
(see HM) is not observed, indicating the presence of additional sources
of ionizing radiation, e.g., QSOs undetected in present surveys
or obscured by intervening dust, or young stars in forming galaxies.

\acknowledgements
MR is grateful to NASA for support through grant HF-01075.01-94A from
the Space Telescope Science Institute. WLWS and TAB were supported by
grant AST92-21365 from the National Science Foundation. 
JM acknowledges support from the W. M. Keck Foundation and from NASA
grant NAG-51618 during his stay at IAS.
DW acknowledges support from NASA Astrophysical Theory Grants
NAG5-2882 and NAG5-3111. LH was supported by NSF grant
ASC93-18185 and the Presidential Faculty Fellows Program.
 We thank the
W.M. Keck Observatory Staff for their help with the observations.
Supercomputing support was provided by the San Diego Supercomputing
Center and the Pittsburgh Supercomputing Center.

\pagebreak

\vfill\eject
\pagebreak

\end{document}